# The EUROnu Project

T.R.Edgecock, O.Caretta, T.Davenne, C.Densam, M.Fitton, D.Kelliher, P.Loveridge,
S.Machida, C.Prior, C.Rogers, M.Rooney, J.Thomason and D.Wilcox
*STFC Rutherford Appleton Laboratory, Didcot, Oxon, OX11 0QX, UK*

E.Wildner, I.Efthymiopoulos, R.Garoby, S.Gilardoni, C.Hansen, E.Benedetto, E.Jensen,
A.Kosmicki, M.Martini, J.Osborne, G.Prior, T.Stora, T.Melo Mendonca, V.Vlachoudis and
C.Waaijer
*CERN, CH-1211 Geneva 23, Switzerland*

P.Cupial
*AGH University of Science and Technology, Krakow, Poland*

A.Chancé, A.Longhin, J.Payet and M.Zito
*Irfu, CEA-Saclay, 91191 Gif-sur-Yvette, France*

E.Baussan, C.Bobeth, E.Bouquerel, M.Dracos, G.Gaudiot, B.Lepers, F.Osswald, P.Poussot,
N.Vassilopoulos, J.Wurtz and V.Zeter
*IPHC, Université de Strasbourg, CNRS/IN2P3, F-67037 Strasbourg, France*

J.Bielski, M.Kozien, L.Lacny, B.Skoczen, B.Szybinski, A.Ustrycka and A.Wroblewski
*Cracow University of Technology, Warszawska 24 St., 31-155 Krakow, Poland*

M.Marie-Jeanne, P.Balint, C.Fourel, J.Giraud, J.Jacob, T.Lamy, L.Latrasse, P.Sortais and
T.Thuillier
*Laboratoire de Physique Subatomique et de Cosmologie,*
*Université Joseph Fourier Grenoble, CNRS/IN2P3,*
*Institut National Polytechnique de Grenoble, Grenoble, France*

S.Mitrofanov, M.Loiselet, Th. Keutgen and Th. Delbar
*Université catholique de Louvain, Centre de Recherche du Cyclotron, Louvain-la-Neuve, Belgium*

F.Debray, C.Trophine, S.Veys and C.Daversin
*Laboratoire National de Champs Magnétiques Intenses, CNRS, Université Joseph Fourier, Grenoble,
France*

V.Zorin, I.Izotov and V.Skalyga
*Institute of Applied Physics, Nizhny Novgorod, Russia*

G.Burt and A.C.Dexter
*Lancaster University, Lancaster, UK*

V.L.Kravchuk, T.Marchi, M.Cinausero, F.Gramegna, G.De Angelis and G.Prete
*INFN, Laboratori Nazionali di Legnaro, Legnaro, Italy*

G.Collazuol, M.Laveder, M.Mazzocco, M.Mezzetto and C.Signorini

*Physics Department, Padova University and INFN, Padova, Italy*

E.Vardaci, A.Di Nitto, A.Brondi, G.La Rana, P.Migliozzi, R.Moro and V.Palladino
*Dipartimento di Scienze Fisiche dell'Universitá di Napoli and INFN, Napoli, Italy*

N.Gelli
*INFN, Sezione di Firenze, Sesto Fiorentino, Firenze, Italy*

D.Berkovits, M.Hass and T.Y.Hirsh,
*Weizmann Institute of Science, Rehovot, Israel and Soreq NRC, Yavne, Israel*

M.Schaumann, A.Stahl and J.Wehner
*Aachen University, Germany*

A.Bross, J.Kopp[1], D.Neuffer and R.Wands
*Fermi National Accelerator Laboratory, Batavia, IL, USA*

R.Bayes, A.Laing and P.Soler
*School of Physics and Astronomy, University of Glasgow, Glasgow, UK*

S.K.Agarwalla, A.Cervera Villanueva, A.Donini[2], T.Ghosh, J.J.Gómez Cadenas,
P.Hernández, J.Martín-Albo and O.Mena
*IFIC, CSIC and Universidad de Valencia, Valencia, Spain*

J.Burguet-Castell
*Universitat de Illes Balears, Spain*

L.Agostino, M.Buizza-Avanzini, M.Marafini, T.Patzak and A.Tonazzo
*APC, Université Paris Diderot, CNRS/IN2P3, CEA/Irfu, Obs. de Paris, Sorbonne Paris Cité, F-75205 Paris France*

D.Duchesneau
*LAPP, Université de Savoie, CNRS/IN2P3, F-74941 Annecy-le-Vieux, France*

L.Mosca
*Laboratoire Souterrain de Modane, F-73500 Modane, France*

M.Bogomilov, Y.Karadzhov, R.Matev and R.Tsenov
*Department of Atomic Physics, St. Kliment Ohridski University of Sofia, Sofia, Bulgaria*

E.Akhmedov, M.Blennow, M.Lindner and T.Schwetz
*Max Planck Institut für Kernphysik, Heidelberg, Germany*

E.Fernández Martinez, M.Maltoni and J.Menéndez
*Departamento de Física Teórica and Instituto de Física Teórica, Universidad Autónoma de Madrid, Madrid, Spain*

---

[1] Also at Max Planck Institut für Kernphysik, Heidelberg, Germany
[2] Also at IFT, CSIC and Universidad Autónoma de Madrid, Madrid, Spain

C.Giunti
*INFN, Sezione di Torino, Torino, Italy*

M.C.González García[3] and J.Salvado
*Departament d'Estructura i Constituents de la Matéria and Institut de Ciencies del Cosmos, Universitat de Barcelona, Barcelona, Spain*

P.Coloma and P.Huber
*Physics Department, Virginia Polytechnic Institute and State University, Blacksburg, VA, USA*

T.Li, J.López Pavón, C.Orme and S.Pascoli
*Institute for Particle Physics Phenomenology, Physics Department, University of Durham, Durham, DH1 3LE, UK*

D.Meloni, J.Tang and W.Winter
*Institut für Theoretische Physik und Astrophysik, Universität Würzburg, Würzburg, Germany*

T.Ohlsson and H.Zhang
*Department of Theoretical Physics, Royal Institute of Technology (KTH) AlbaNova University Centre, Stockholm, Sweden*

L.Scotto-Lavina
*Physik-Institut, University of Zurich, CH-8057 Zurich, Switzerland.*

F.Terranova
*INFN, Laboratori Nazionali di Frascati, Frascati, Italy*

M.Bonesini
*INFN, Sezione Milano Bicocca, Milano, Italy*

L.Tortora
*INFN, Sezione Roma 3, Roma, Italy*

A.Alekou, M.Aslaninejad, C.Bontoiu, A.Kurup, L.J.Jenner[4], K.Long, J.Pasternak and J.Pozimski
*Imperial College, London, SW7 2BW, UK*

J.J.Back and P.Harrison
*University of Warwick, Coventry, CV4 7AL, UK*

K.Beard
*Muons Inc., Batavia, USA*

A.Bogacz
*Thomas Jefferson National Accelerator Facility, Newport News, Virginia, USA*

---

[3] Also at Institució Catalana de Recerca i Estudis Avançats (ICREA) and C.N.Yang Institute for Theoretical Physics, State University of New York at Stony Brook, Stony Brook, USA

[4] Also at FNAL, Batavia, USA

J.S.Berg, D.Stratakis and H.Witte
*Brookhaven National Laboratory, Upton, New York, USA.*

P.Snopok
*Illinois Institute of Technology, Chicago, USA*

N.Bliss, M.Cordwell, A.Moss and S.Pattalwar
*STFC Daresbury Laboratory, Daresbury, WA4 4AD, UK*

M.Apollonio
*Diamond Light Source, Didcot, Oxon, OX11 0QX, UK*

**Abstract**
The EUROnu project has studied three possible options for future, high intensity neutrino oscillation facilities in Europe. The first is a Super Beam, in which the neutrinos come from the decay of pions created by bombarding targets with a 4 MW proton beam from the CERN High Power Superconducting Proton Linac. The far detector for this facility is the 500 kt MEMPHYS water Cherenkov, located in the Fréjus tunnel. The second facility is the Neutrino Factory, in which the neutrinos come from the decay of $\mu^+$ and $\mu^-$ beams in a storage ring. The far detector in this case is a 100 kt Magnetised Iron Neutrino Detector at a baseline of 2000 km. The third option is a Beta Beam, in which the neutrinos come from the decay of beta emitting isotopes, in particular $^6$He and $^{18}$Ne, also stored in a ring. The far detector is also the MEMPHYS detector in the Fréjus tunnel. EUROnu has undertaken conceptual designs of these facilities and studied the performance of the detectors. Based on this, it has determined the physics reach of each facility, in particular for the measurement of CP violation in the lepton sector, and estimated the cost of construction. These have demonstrated that the best facility to build is the Neutrino Factory. However, if a powerful proton driver is constructed for another purpose or if the MEMPHYS detector is built for astroparticle physics, the Super Beam also becomes very attractive.

## Introduction

The discovery that the neutrino changes type, or flavour, as it travels through space, a phenomenon referred to as neutrino oscillations, implies that neutrinos have a tiny, but non-zero mass, that lepton flavour is not conserved, and that the Standard Model of particle physics is incomplete [1]. The implications of these observations are far reaching: neutrino interactions may be responsible for the removal of the anti-matter created in the Big Bang from the early Universe; the neutrino may have played a central role both in creating the homogeneous Universe in which we live and in the formation of the galaxies; and, perhaps most fundamental of all, the neutrino may have played a crucial role in the birth of the Universe itself. Knowledge of the contribution of neutrinos in these areas requires precise measurements of parameters governing neutrino oscillations.

The theoretical description of neutrino oscillations is based on the assumption that there are three neutrinos, each of which has a tiny mass (the mass eigenstates). No two neutrinos have the same mass. Under this assumption, quantum mechanics implies that the three neutrino flavours may be considered to be mixtures of the three mass eigenstates, the relative weight of the mass eigenstates differing from one neutrino flavour to another. The standard phenomenological description of neutrino oscillations then has four mixing parameters and two mass-difference parameters. Three of the mixing parameters take the form of mixing angles, labeled $\theta_{12}$, $\theta_{23}$, and $\theta_{13}$, while the fourth is a phase parameter, labeled $\delta_{CP}$, which, if it is non-zero, causes the interactions of neutrinos to be different to those of anti-neutrinos, violating the matter-antimatter symmetry that is present in the Standard Model i.e. CP-violation. The mass differences are given by $\Delta m_{12}^2 = m_2^2 - m_1^2$ and $\Delta m_{23}^2 = m_3^2 - m_2^2$, where $m_1$, $m_2$ and $m_3$ are the masses of the mass eigenstates.

At the start of EUROnu, only the first two mixing angles and the two mass differences had been measured. It was already clear that to measure the remaining parameters, the last mixing parameter, $\theta_{13}$, the CP phase and the sign of $\Delta m_{23}^2$ (the so-called mass hierarchy), would require new neutrino oscillation facilities. During EUROnu, a number of new facilities, in particular three in which neutrinos are produced in nuclear reactors (Daya Bay, Double Chooz and Reno) and one which makes them using a proton accelerator (T2K), have made the first measurement of the angle $\theta_{13}$. These have demonstrated that this angle is large, around 9° [2, 3, 4, 5], which means that the remaining two unknown parameters of neutrino oscillation are now within reach and that precise measurements are possible. However, a new, high intensity neutrino oscillation facility with better controlled systematic errors is required to deliver these physics goals.

EUROnu was a Design Study within the European Commission Seventh Framework Programme, Research Infrastructures. It has investigated three possible options for the future, high intensity neutrino oscillation facility in Europe able to make these measurements. The work was done by the EUROnu consortium, consisting of 15 partners and a further 15 associate partners [6].

The three facilities studied are:
- The CERN to Fréjus Super Beam, using the 4 MW version of the Superconducting Proton Linac (SPL) at CERN [7]. The baseline far detector is a 500 kT fiducial mass water Cherenkov detector, MEMPHYS [8].

- The Neutrino Factory, in which the neutrino beams are produced from the decay of muons in a storage ring. This work has been done in close collaboration with the International Design Study for a Neutrino Factory (IDS-NF) [9].
- The Beta Beam, in which the neutrino beams are produced from the decay of beta emitting ions, again stored in a storage ring.

The project started on 1st September 2008 and ran for four years. The work done on the accelerator facilities, the detectors and on determining the physics performance and the cost are described in the following sections.

**The Super Beam**

A Super Beam creates neutrinos by impinging a high power proton beam onto a target and focussing the pions produced towards a far detector using a magnetic horn. The neutrinos come from the decay of pions in a decay tunnel following the target, thus producing a beam in the direction of the tunnel (see Figure 1). EUROnu is studying the CERN to Fréjus Super Beam, using the High Power Superconducting Proton Linac (HP-SPL) [7] as the proton driver, producing a 4 MW beam. Full details of the work done can be found in [10]. The baseline is 130 km and the planned far detector is the 500 kT fiducial mass MEMPHYS water Cherenkov detector [8]. This would be built in two new caverns in the Fréjus tunnel.

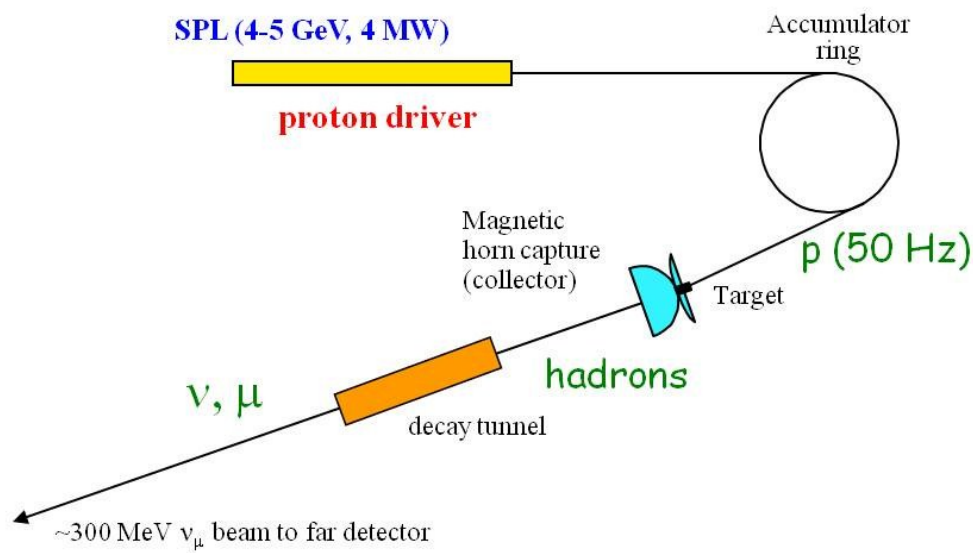

*Figure 1: Layout of the CERN to Fréjus Super Beam.*

The High Power Superconducting Proton Linac will produce a 4 MW beam at 5 GeV and operate at a frequency of 50 Hz. It will consist of a number of sections. The first is these, up to 160 MeV, will be about 90 m long and will be normally conducting. The low power version of this, Linac 4 [12], is currently under construction at CERN. The remaining three sections will be superconducting and will accelerate the beam to 0.7, 2.5 and 5 GeV, respectively. The SPL will accelerate 42 bunches in a pulse, with a pulse duration of 600 µs, which is too long for Super Beam operation. Hence an Accumulator ring will be used to reduce the number of bunches to 6, each 120 ns in length with gaps of 60 ns, resulting in a pulse length less than the 5 µs requirement coming from horn operation. A significant amount of design work has already been done on the SPL and R&D has started on many components [13].

Given the difficulty in producing a single target and horn able to work in a 4 MW beam, the option taken in EUROnu is to use four of each instead. The beam from the Accumulator will then be steered on to each target in turn, so that they all run at 12.5 Hz rather than 50 Hz and receive 1 MW. For the targets and the horns, this results in a smaller extrapolation from technology already in use. To achieve this, a system of two kicker and four bending magnets has been designed to steer the beam on to each target in turn (see Figure 2).

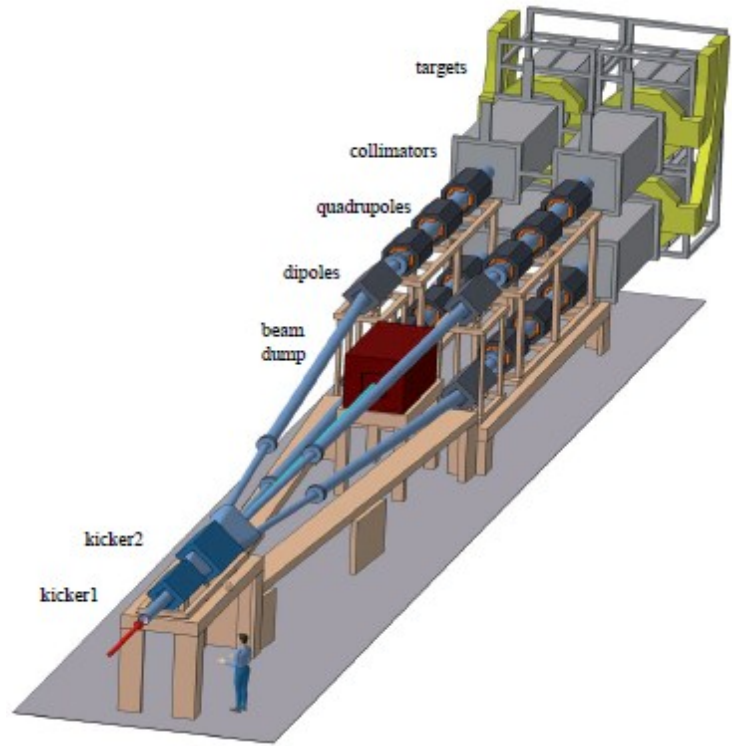

*Figure 2: The Super-Beam beam transport and distribution system.*

An outline design for the 4 target and horn system is shown in Figure 3. To minimize the production of thermal neutrons and hence reduce the heat load and radiation damage to the surrounding horns, the baseline design for the target is a pebble bed, consisting of 3 mm diameter spheres of titanium in a canister, 200 mm long (see Figure 4). These are cooled by flowing helium gas through vents in the canister, at around 10 bar pressure. Thermal modeling shows that this should be sufficient to cool the targets up to a few MW. To verify this, offline tests of the cooling system will be undertaken in the future. These will use an inductive coil to heat the target at the required level and demonstrate that this heat can be successfully removed. A test target will also be subjected to a beam of the correct energy density using the HiRadMat [14] facility at CERN, to further verify the cooling and demonstrate that the titanium spheres and the target structure can withstand the thermal shock from the beam.

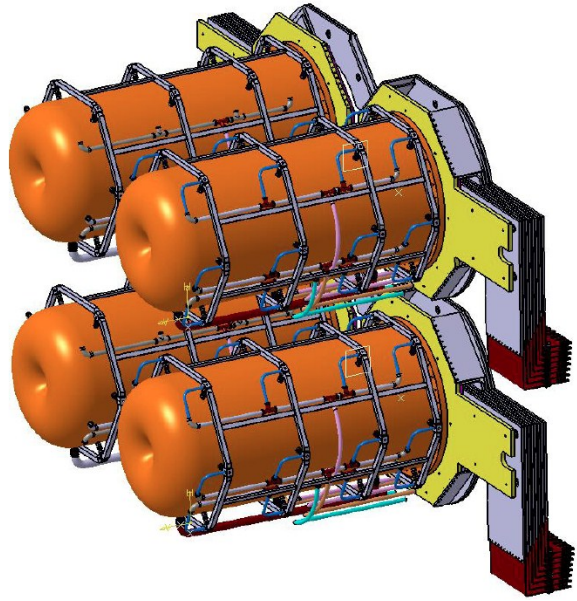

*Figure 3: Conceptual engineering design of the 4 target and horn system for the Super Beam.*

The focusing horn design has been optimized for the CERN to Fréjus beam. It will employ a single horn around the target, and will not have a reflector. As for the targets, four horns will be used and will need to be pulsed at 350 kA, resulting in significant heating. Further heating will come from beam loss, resulting in a maximum of 12 kW on the surface around the target. Studies with thermal codes show that this can be removed with water cooling of the outer surface of the horn. The thermal stresses in the horn material resulting from the heating are a maximum of 18 MPa and prototype tests will be required to determine what the lifetime of a horn will be due to the resulting fatigue and from radiation damage. A support system for the 4 horn system under this load has been designed. The final aspect of the horn system is a pulsing circuit to deliver the required current at up 17 Hz (in case of the failure of one target+horn combination). A circuit to do this has been designed and it is planned to build a prototype of it (see Figure 5).

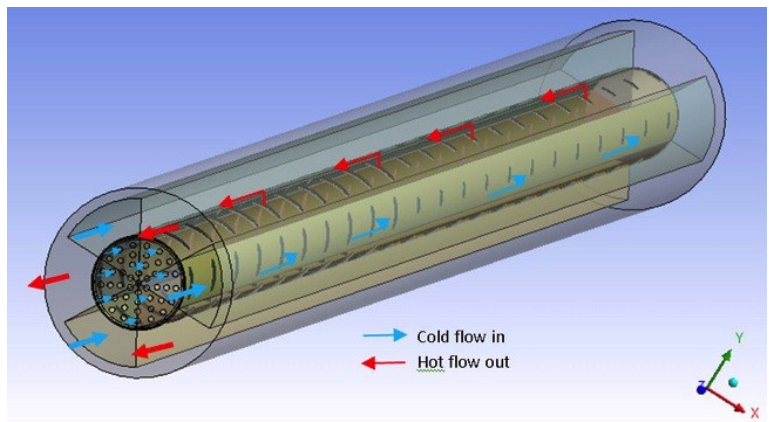

*Figure 4: Proposed pebble bed target for the Super Beam.*

The targets and horns will need to be mounted in a target station which allows the change, storage and maintenance of targets and horns, in case of failure. To enable this, the target station will have a number of separate sections and activation studies have been done to determine the shielding requirements for each. A design of this has been made, based on these studies and experience gained with the T2K target station. It incorporates remote handling facilities, a hot cell for maintenance and a storage area for old targets and horns, called the morgue. It will allow access to the critical components of the system, for example the power supplies for the horns, and will allow the safe removal of activated components for disposal. The section of the target station that contains the targets and horns is shown in Figure 6.

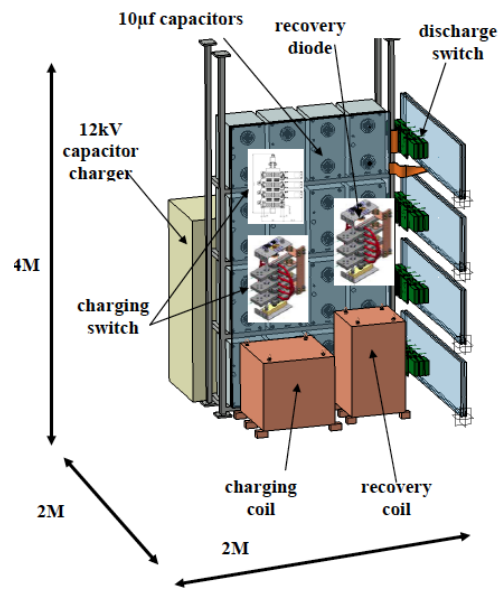

*Figure 5: Design of the pulsing system for the Super Beam horns.*

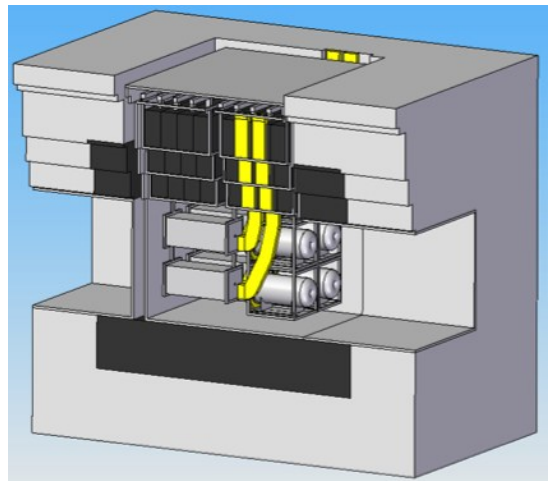

*Figure 6: Conceptual design of the section of the target station for the 4 targets and horn systems. The beam enters from the left. The horn pulsing circuits will be mounted on top of the shielding, so the strip lines exit vertically.*

To allow the determination of the physics performance, the pion and neutrino production by the Super Beam have been simulated. The resulting flavour composition of the beam is shown in Figure 7 for both neutrino and anti-neutrino beams. Note that the $\nu_e$ contamination in the beam is significantly less than 1% in both cases.

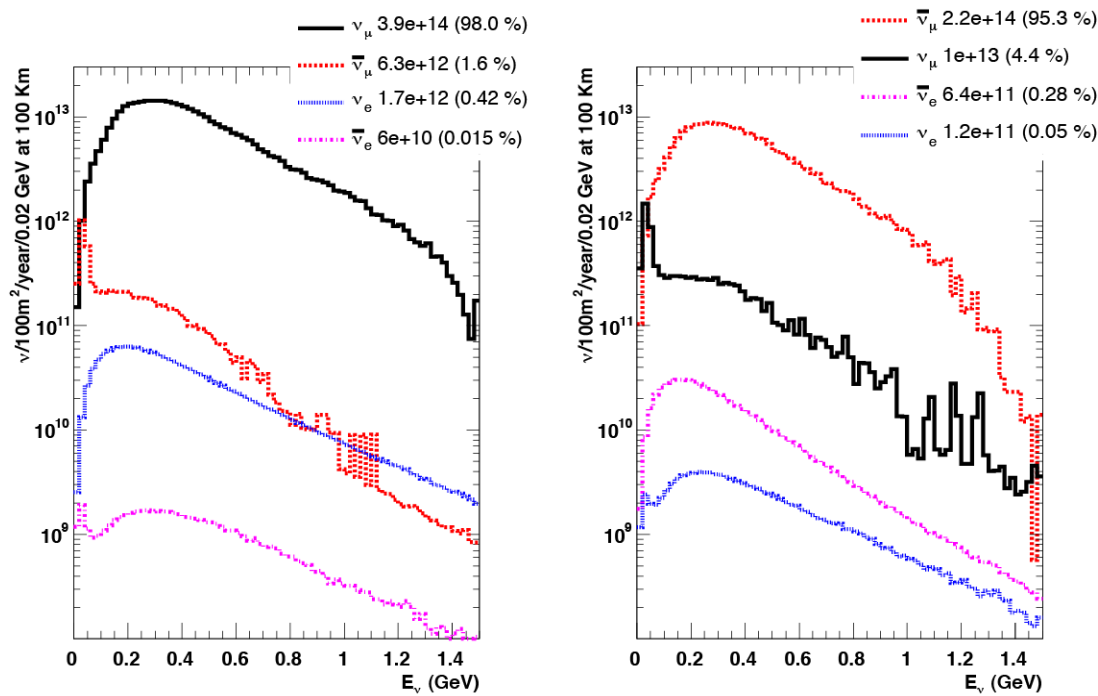

*Figure 7: The composition of the neutrino beam produced by the Super Beam facility.*

### Neutrino Factory

In a Neutrino Factory [15], the neutrinos are produced from the decay of muons in a storage ring. The muons are produced by impinging a 4 MW proton beam onto a heavy metal target and focussing the pions produced into a decay channel using a 20 T super-conducting solenoid. In the original baseline, the muons from the pion decay are captured, bunched, phase rotated and finally cooled in the muon front-end, before being accelerated using a linac, two re-circulating linear accelerators (RLAs) and a non-scaling Fixed Field Alternating Gradient accelerator (ns-FFAG) to 0.9 GeV, 3.6 GeV, 12.6 GeV and 25 GeV, respectively (see Figure 8). The muons are then injected into two storage rings, to produce beams of neutrinos and anti-neutrinos to two far detectors. Stored $\mu^+$ beams will produce pure electron neutrino and muon anti-neutrino beams, while $\mu^-$ will produce pure electron anti-neutrino and muon neutrino beams. To be able to distinguish signal from background, it is essential that the far detector can separate $\mu^+$ from $\mu^-$ with high efficiency. As a result, the baseline detector is a Magnetised Iron Neutrino Detector (MIND).

However, following the recent measurement of $\theta_{13}$, the required muon energy has been reduced to 10 GeV and only one decay ring will be used. The envisaged neutrino baseline is now around 2000 km.

Two options are under consideration for the Neutrino Factory proton driver. The first is a super-conducting linear accelerator. Indeed, if the facility was to be built at CERN, this would be the HP-SPL [7]. This would be followed by an Accumulator ring, as for the Super Beam, and a Compressor ring to reduce the proton bunch length to 3 ns. The other option employs a rapidly cycling synchrotron, working at 50 Hz, to accelerate the beam to 10 GeV. This would use a normally conducting linear injector to accelerate the beam to 180 MeV.

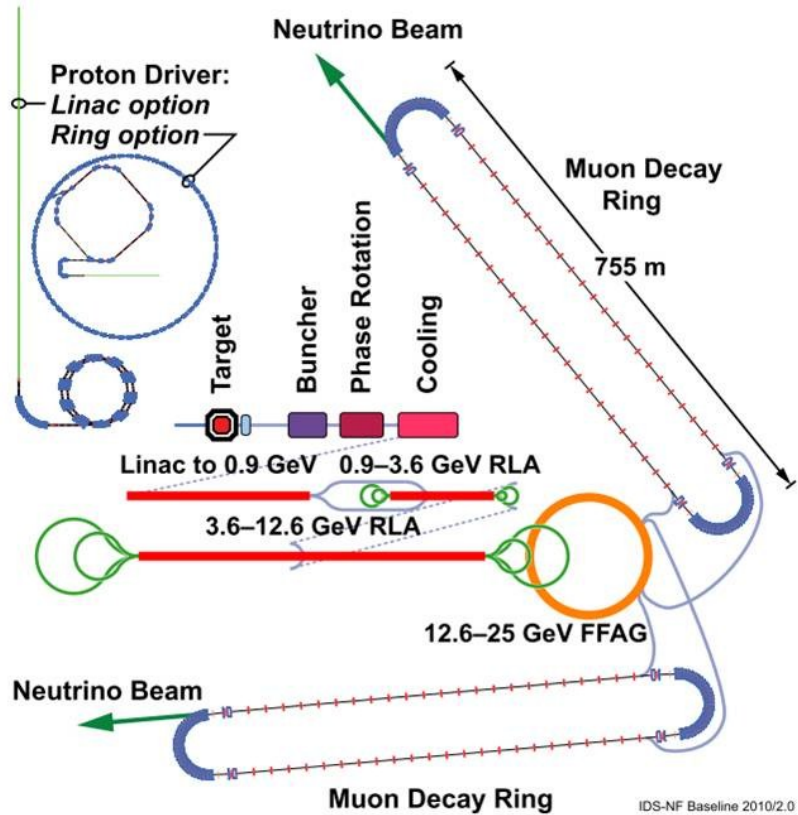

*Figure 8: Original baseline layout of the Neutrino Factory.*

The baseline pion production target is a continuous liquid mercury jet. This would be fired across the proton beam at a small angle so that the beam and target overlap for two interaction lengths. The pions produced would be focused by a combined normal and super-conducting magnet of 20 T around the target (see Figure 9). Both the beam and target would also be at a small angle to the axis of the solenoidal field, so that the mercury collects in a pool. As well as allowing the mercury to be re-circulated, this could also form a part of the proton beam dump. The magnetic field would be ramped down adiabatically to 1.5 T at the entrance of the pion decay channel, using a succession of superconducting coils. However, simulations of secondary particle production in the target and subsequent absorption in the super-conducting coils have shown that the heat load in the coils around and close to the target is much too large, up to 50 kW. The main problem comes from secondary neutrons.

Various options are being considered to reduce this heating. The most obvious is simply to add more shielding. It has been demonstrated that this will work, but it would mean that the radius of the super-conducting coils would double, making these significantly more difficult to build and operate. A study of pion production has shown that similar production rates to those in mercury can be achieved with lower atomic number elements (see Figure 10), but these may produce significantly fewer neutrons. As a result, targets with lower atomic number are under study. An interesting candidate is gallium, which has a low enough melting point that it could be used as a liquid, in a similar way to a mercury jet. In addition, the fact that it is a solid at room temperature makes storage and disposal after activation significantly easier.

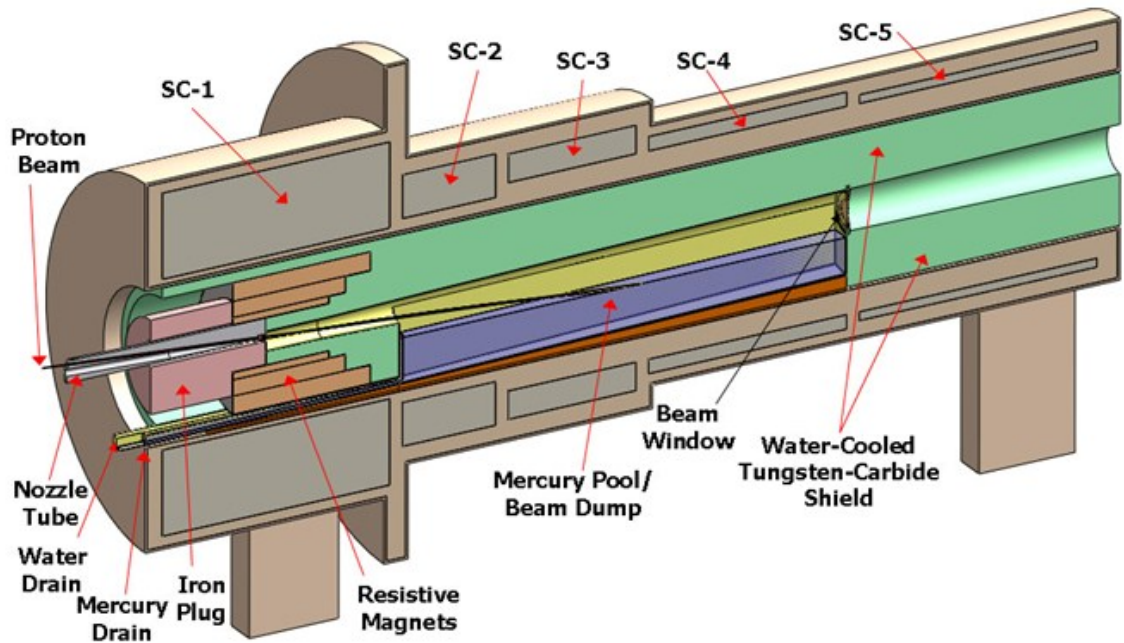

*Figure 9: Conceptual layout of the Neutrino Factory pion production target and capture system.*

The target is followed by the pion decay channel and the muon front-end. The former is a solenoidal channel of 100 m length, employing 1.5 T magnets to maximize the captured muon flux. The aim of the muon front-end is to prepare the muon beam for acceleration. It consists of a chicane, a buncher, a phase rotator and a cooling channel. The chicane is required because as well as the required large flux of muons in the front-end, there are also still many protons, pions and electrons. If nothing is done about these, they will be lost throughout the front-end, resulting in levels of activation about 100 times above the canonical level for hands-on maintenance. As a result, the chicane is used to remove the higher momentum unwanted particles. It is followed by an absorber, to remove those at lower momentum. The efficiency for transmission of useful muons is about 90%, while the unwanted particles are reduced to a manageable level. The chicane is followed by a section, 33 m long, which uses RF cavities to bunch the beam. This in turn is followed by a phase rotation section 42 m long, which utilises the correlation between position in the bunch train and energy that has built up by this stage. It uses RF cavities to slow down the faster going muons at the front and speed up the slower going particles at the back and thereby reduces the energy spread of the beam.

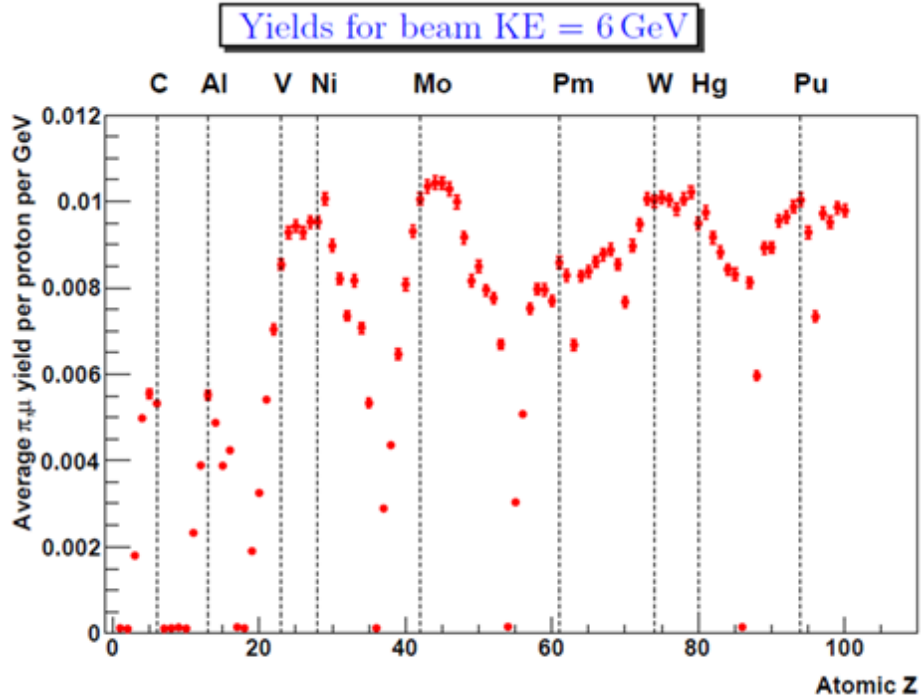

*Figure 10: Pion production as a function of atomic number, assuming a cylindrical target 20cm long and 2cm in diameter.*

The phase rotation section is followed by the cooling channel, which will employ the technique of ionization cooling. In this, an absorber is used to reduce both the longitudinal and transverse components of the muon momentum. The lost longitudinal momentum is then restored using RF cavities, giving a net reduction in transverse momentum and hence transverse cooling. However, as well as cooling through energy loss, the absorber also heats through multiple scattering and the best balance between the two is achieved by using a low atomic number material, such as liquid hydrogen or lithium hydride. In addition, the cooling efficiency is significantly increased if the absorber is in region in which the beam is highly convergent or divergent, thus requiring a superconducting field around the absorber region. Superconducting magnets are also required around the RF cavities to aid transport. The result is that the cooling cell is a complex object (see Figure 11).

Due to the complexity, an engineering demonstration of the cooling technique is being constructed at the STFC Rutherford Appleton Laboratory. This project, called MICE [16], is due to give a first demonstration of ionisation cooling during 2013. In addition, as the RF cavities of the baseline cooling cell will be in a large magnetic field, measurements of the effect this will have on the accelerating gradient are being made by the MuCool project [17]. To minimize potential problems, alternative cooling lattices are being studied that reduce the magnetic field at the cavities, while maintaining the same performance [18].

Following the reduction of muon energy to 10 GeV, two options now exist for the muon acceleration system (see Figure 12). The first uses a linac to 0.8 GeV, followed by two Re-circulating Linear Accelerators (RLAs), one to 2.8 GeV and the second to 10 GeV. The second option uses a linac to 1.2 GeV, an RLA to 5 GeV and a non-scaling Fixed Field

Alternating Gradient (ns-FFAG) accelerator to 10 GeV. Both options are under study to determine which would be best based on performance and cost.

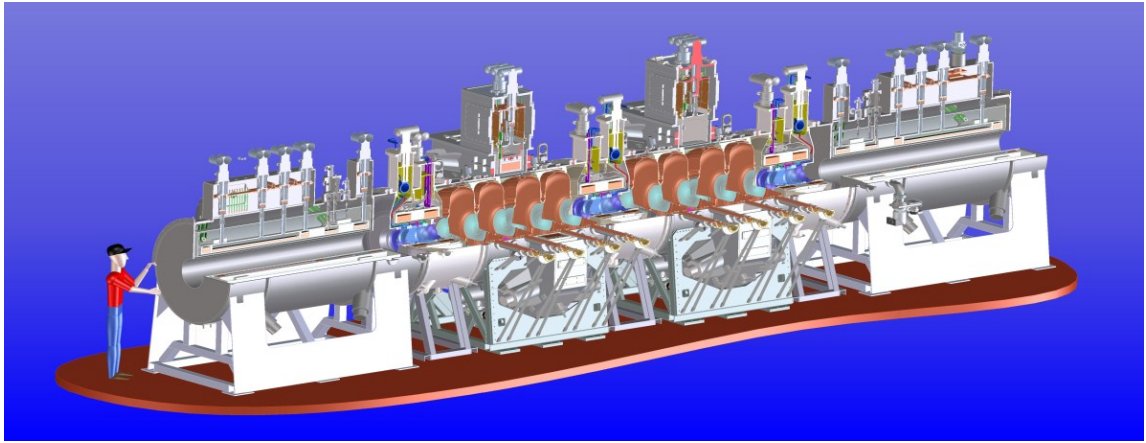

*Figure 11: Engineering drawing of the MICE experiment [16]. The central region shows two ionization cooling cells, with instrumentation regions on either side for measuring the parameters of muons entering and leaving these cells.*

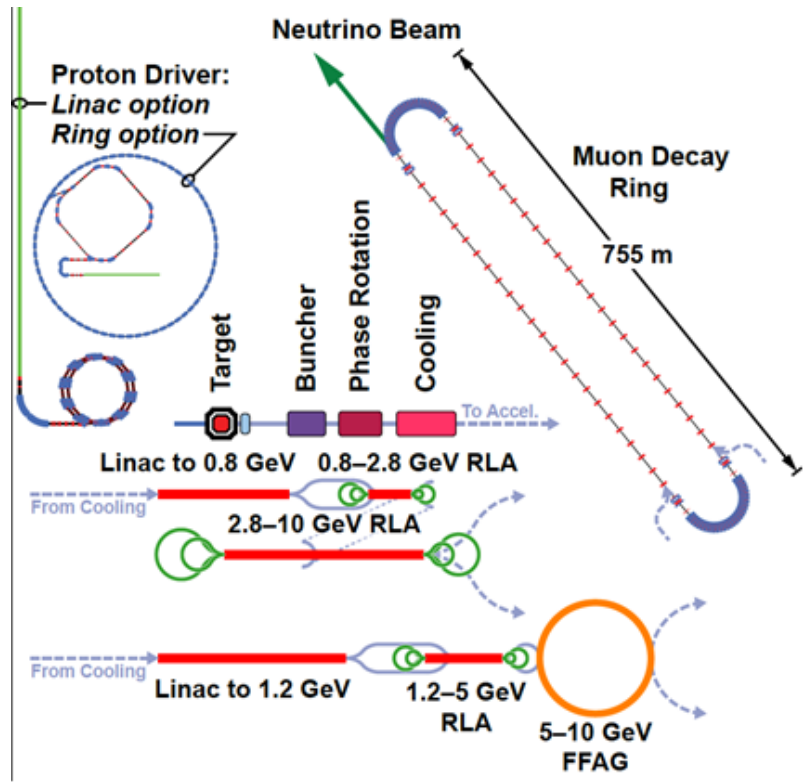

*Figure 12: Updated layout for a Neutrino Factory for 10 GeV operation, with only one decay ring. The two options for muon acceleration are shown.*

An ns-FFAG is proposed as its properties of fixed magnetic fields and pseudo-isochronous operation mean that muon acceleration will be very fast, plus it has the large acceptance required for the high emittance muon beam, even after cooling. However, it is an entirely novel type of accelerator, so a proof-of-principle machine called EMMA [19] has been constructed at the STFC Daresbury Laboratory (see Figure 13). This has recently demonstrated that many of the novel features of the muon accelerator, in particular serpentine acceleration and multiple resonance crossings [20], work. The full EMMA experimental programme has started and will study the remaining issues.

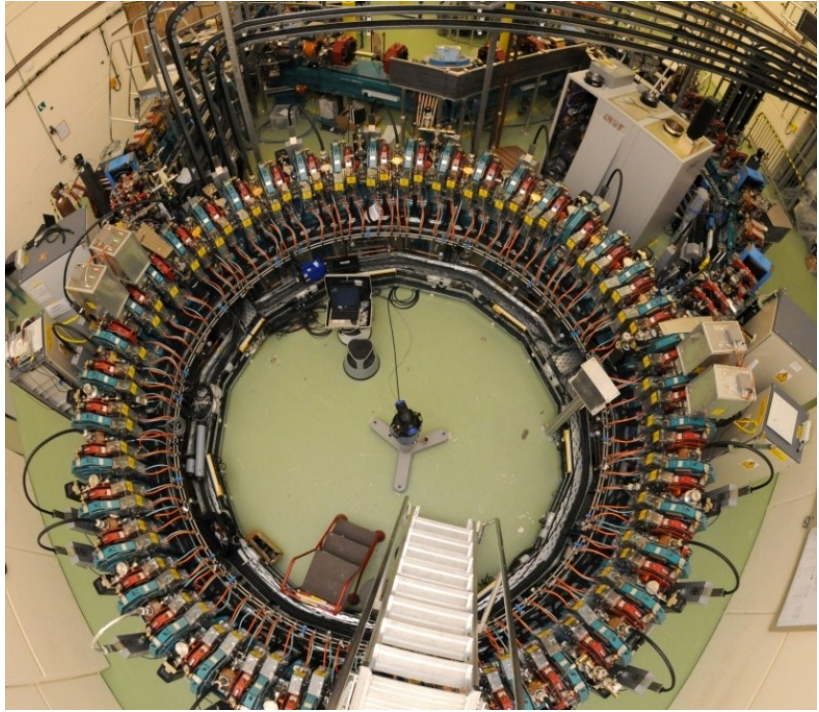

*Figure 13: The EMMA proof-of-principle accelerator at the Daresbury Laboratory.*

The final part of the Neutrino Factory is the decay ring. It is planned to produce and accelerate bunches of $\mu^+$ and $\mu^-$ at the same time. These will arrive in three bunches each, of 250 ns length, separated by 120 μs. The decay ring will have a total circumference of 1286 m, of which 470 m will form a production straight for neutrinos in the direction of the far detector for both muon charges. The ring will be tilted at an angle of about $10°$ degrees for the 2000 km long baseline. An outline injection system design has been made that will inject all of the bunches into the ring. A minimum separation of at least 100 ns is required between bunches to make it possible to determine which bunch detected neutrinos come from. With the expected 2% energy spread of the muon beam, this will exist for 4 muon lifetimes, allowing the vast majority of muons to decay.

**The Beta Beam**

The production of (anti-)neutrinos from the beta decay of radioactive isotopes circulating in a race track shaped storage ring was proposed in 2002 [21]. Beta Beams produce pure beams of electron neutrinos or antineutrinos, depending on whether the accelerated isotope is a $\beta^+$ or a $\beta^-$ emitter. The facility discussed here is based on CERN's infrastructure and will re-use some existing accelerators, though with modifications. This will significantly reduce the cost compared to a green field site, though it will constrain the performance (see Figure 14). It will consist of an ion production system, using a proton driver to accelerate particles and create the required ion species in a target. This will be followed by an ion collection device and a 60 GHz ECR source for bunching. There will then be an ion acceleration system, using a linac to 100 MeV, a Rapid Cycling Synchrotron, the existing Proton Synchrotron and the Super Proton Synchrotron, before injection into a decay ring [22].

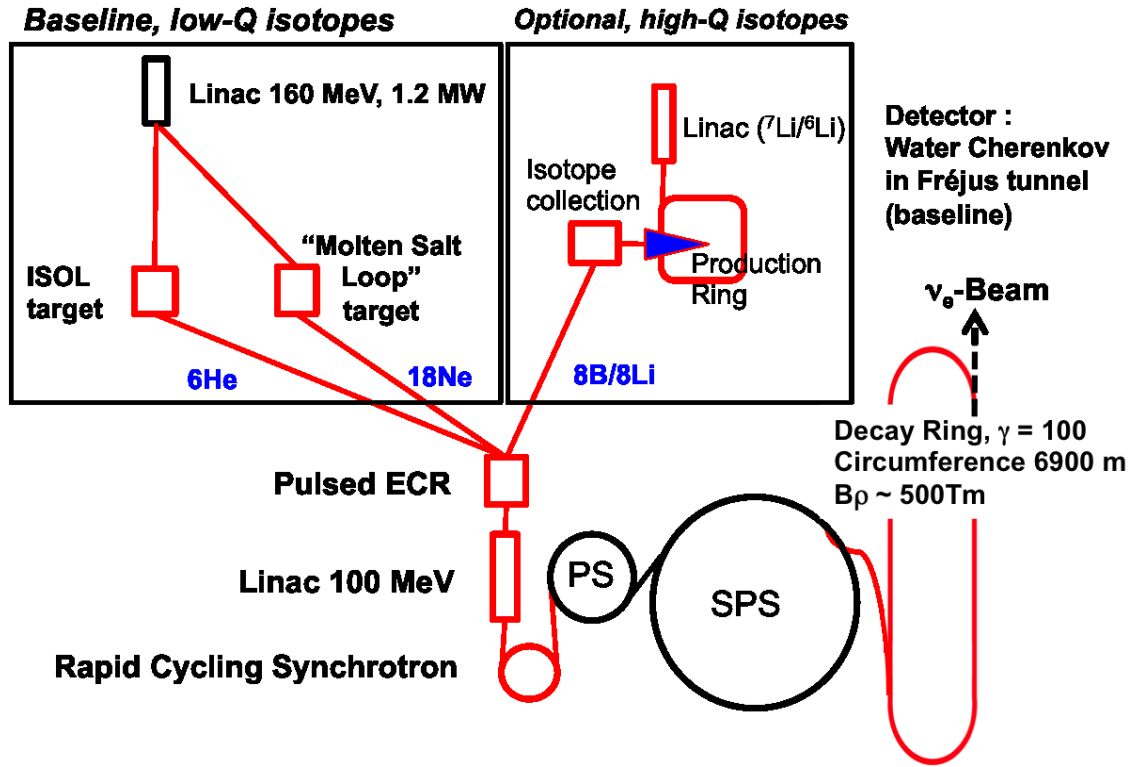

*Figure 14: Layout of the CERN Beta Beam.*

One of the most important issues for a Beta Beam is the production, acceleration and storage of a sufficient flux of ions to meet the physics goals. The isotope pair that was first studied for neutrino production, in the EURISOL FP6 Design Study [23], is $^6$He and $^{18}$Ne, accelerated to $\gamma = 100$ in the SPS and stored in a Decay Ring [24]. Physics studies have indicated that the required fluxes of these ions are $6 \times 10^{13}$ and $1 \times 10^{13}$ ions/second, respectively. At the end of EURISOL, it looked possible to produce the required flux of $^6$He, but that of $^{18}$Ne looked a factor of 20 too small. This has subsequently been addressed in two ways. The first was to consider a production ring (12 m circumference) with an internal gas jet target [25] to make an alternative ion pair, $^8$Li and $^8$B. As the neutrinos from the decay of these ions have about 5 times larger energy than those for $^6$He and $^{18}$Ne, the required baseline has to be 5 times larger and the flux of ions required for the same physics is $10^{14}$ ions/second. In the production ring, a 25 MeV beam of $^7$Li and $^6$Li is injected over a gas jet target of d or $^3$He, respectively. To determine the production rate, the double differential cross-sections for both processes, $^7$Li(d,p)$^8$Li and $^6$Li($^3$He,n)$^8$B, have been measured at the Laboratori Nazionali de Legnaro in Italy [26]. The first measurements were performed using the 8πLP experiment (see Figure 15) and are comparable with results obtained at lower energy. The $^8$B production cross-section was measured using Time of

Flight techniques. The results from this are consistent with theoretical calculations, but three times larger than measurements performed using a different technique. This is still being investigated.

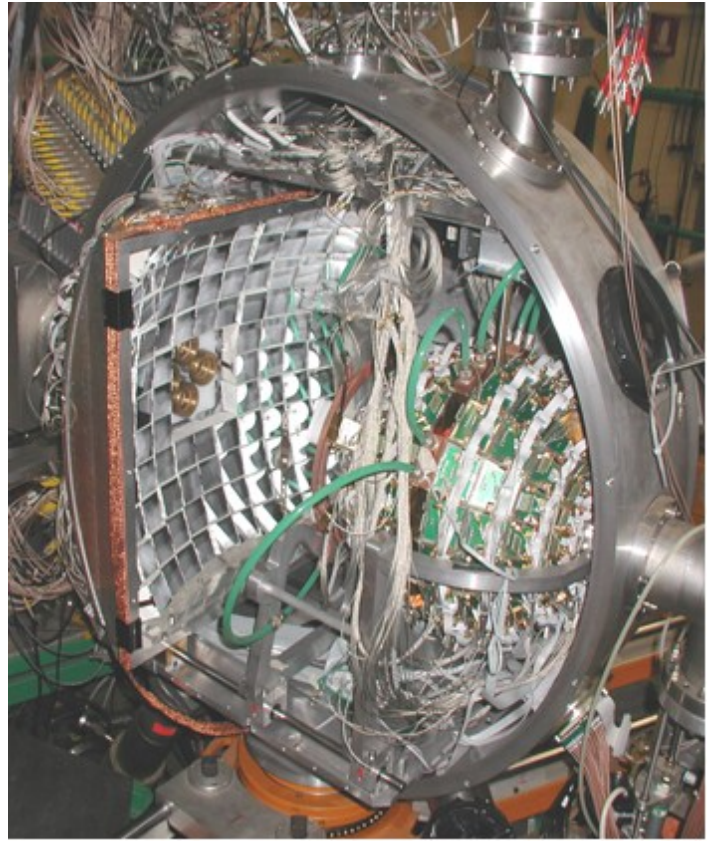

*Figure 15: The 8πLP experiment used for measurement of the $^7Li(d,p)^8Li$ cross-section.*

Based on these measurements, significant design work has been done on the production ring and a prototype device for collection of the ions has been built and tested (see Figure 16). The studies have shown that the thickness of the gas jet target needed to produce the required flux of ions, $10^{19}$ atoms/cm$^2$, is four orders of magnitude bigger than any in current use and will create significant problems for the ring vacuum. Alternative production possibilities have been looked at, for example liquid lithium films, but it remains extremely difficult to meet the ion production goals.

As a result, research on a novel $^{18}$Ne production method, using a molten salt loop (NaF) by the reaction $^{19}F(p,2n)^{18}Ne$, is currently being undertaken (see Figure 17). Modelling suggests that this could achieve the required production rate with a 160 MeV proton linear accelerator at a current of 7 mA. This would be achievable at CERN with an upgrade of Linac 4 [12]. An experiment to validate the method took place at ISOLDE at CERN in June 2012 and demonstrated that the required flux could be achieved [27]. As a result of the work done so far, the $^6$He and $^{18}$Ne ion pair is the recommended baseline for the Beta Beam.

To accept the intense continuous flux of $^6$He or $^{18}$Ne produced, ionize the gas and bunch the ions with the high efficiency, it is planned to use a 60 GHz pulsed Electron Cyclotron

Resonance (ECR) ion source. A prototype device called SEISM (Sixty gigahertz ECR Ion Source using Megawatt magnets) has been designed and the magnetic confinement structure successful built and tested (see Figure 18). It is planned to test plasma production at 28 GHz, to allow comparison with existing ion sources, before proceeding to a 60 GHz plasma.

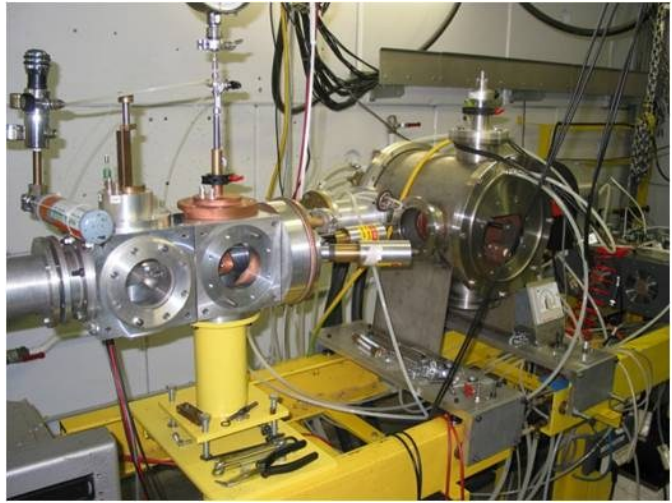

*Figure 16: The prototype ion collection device constructed for Beta Beam studies.*

As shown in Figure 14, after bunching, the ions will be accelerated to 100 MeV/u using a purpose built linear accelerator about 110 m long. This will be followed by a Rapid Cycling Synchrotron, 251 m in circumference, that will accelerate the ions to a maximum magnetic rigidity of 14.47 Tm, corresponding to 3.5 GeV protons, 787 MeV/u for $^6$He$^{2+}$ and 1.65 GeV/u for $^{18}$Ne$^{10+}$. Final acceleration of the ion beams will take place in the existing PS and SPS. Simulations of these show that, although not optimal, they can deliver the required performance. Preliminary activation studies have also been done and these show that the effect of the Beta Beams compared to high intensity proton running varies with the component or material being activated, but the rate is never significantly higher and this should not prevent operation. Collective effects studies have been started. For the SPS results show that the cycling and bunching of the beam has to revisited and optimized for the whole accelerator chain.

The final element of the Beta Beam is the decay ring. As for the Neutrino Factory this will have a race track shape, with a total circumference the same size as the SPS, 6.9 km, and a production straight which is 37% of this size to maximize the neutrino flux. The bunches are injected into the Decay Ring from the SPS for every Beta Beam cycle to compensate for decay losses. The preferred method of doing this is to use a dual frequency RF and inject new beam at a slightly different energy from that already in the ring. The voltage and phase of the two cavity families will then be varied to perform the merging. This technique has been simulated and in part successfully tested. As the ring will use superconducting magnets, the decay losses are a significant problem. The solution is to use coil free mid-plane magnets, so that the deposited power in the magnet coils can be reduced to avoid magnet quench. Another major problem is collective effects and these ultimately will limit the intensity in the ring. In particular, the so-called head-tail effect, in which particles in the tail of the bunch are affected by the field created by the particles in the head, is a serious problem. Although the intensity limit is above the required intensity for $^6$He, this is not true for $^{18}$Ne, where it is only about 20% of that required. Studies are continuing to find a solution to this problem.

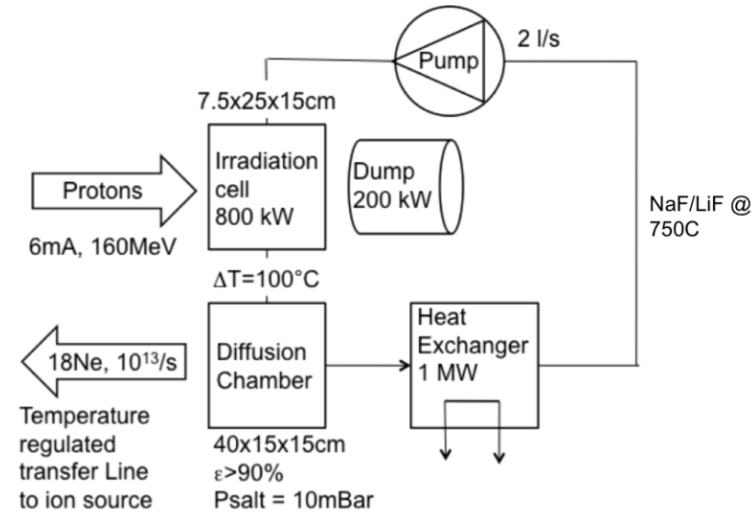

*Figure 17: A NaF molten salt loop for the production of $^{18}$Ne ions.*

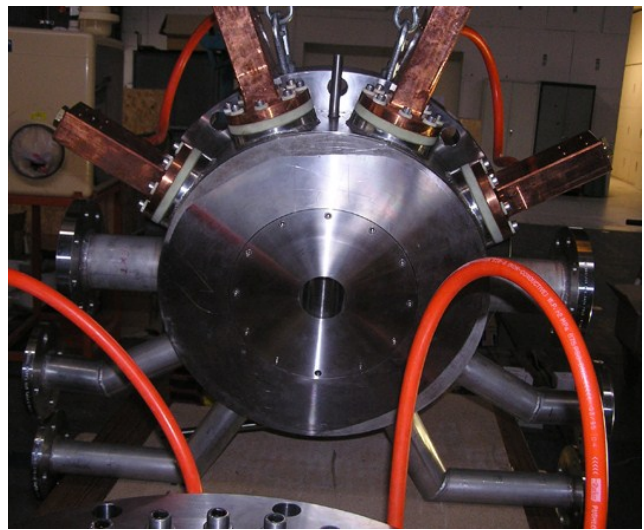

*Figure 18: The SEISM 60 GHz ECR source prototype.*

As a far detector, the baseline isotopes, $^6$He and $^{18}$Ne could use the MEMPHYS detector [8] in the Fréjus tunnel, at a distance of 130 km. Due to the higher energy of the neutrinos, the $^8$Li and $^8$B option would need a detector at some 700 km and may need a different detector technology, such as liquid Argon [28]. Only the first option is considered in this paper.

**Detectors**

The focus of EUROnu has been on the conceptual design of the accelerator facilities. Nevertheless, to make a genuine determination of the physics reach of each facility, it is also important to include the neutrino detectors in the study. Thus, the project has studied the baseline detectors for each facility, with the aim of determining their performance in detecting neutrinos and delivering physics measurements.

The baseline for the Neutrino Factory is a Magnetised Iron Neutrino Detector (MIND) [29]. This is an iron-scintillator calorimeter, with alternating planes of 3 cm thick iron and 2 cm thick solid scintillator. One detector is now planned, of 100 kT mass at a distance of around 2000 km. From CERN, this baseline is possible with a detector in the Pyhäsalmi mine in Finland. The design, shown in Figure 19, has been based on that of the MINOS detector [30]. It will have a transverse size of 14 by 14 m and be 140 m long, meeting the constraints coming from typical underground laboratories [31]. It will have a toroidal magnetic field of >1T to distinguish $\mu^+$ and $\mu^-$ events. Detailed simulations of the detector performance have made using GENIE to generate the neutrino events and GEANT 4 for the detector modeling. Events are reconstructed using, for example, a Kalman filter for track reconstruction. Some results are shown in Figure 20.

Migration matrices, which relate the true neutrino energy to the reconstructed energy, have been produced for MIND, for use in the physics reach determinations. In addition, the systematic errors on the reconstruction of signal and background events have been conservatively estimated at 2% and 5%, respectively.

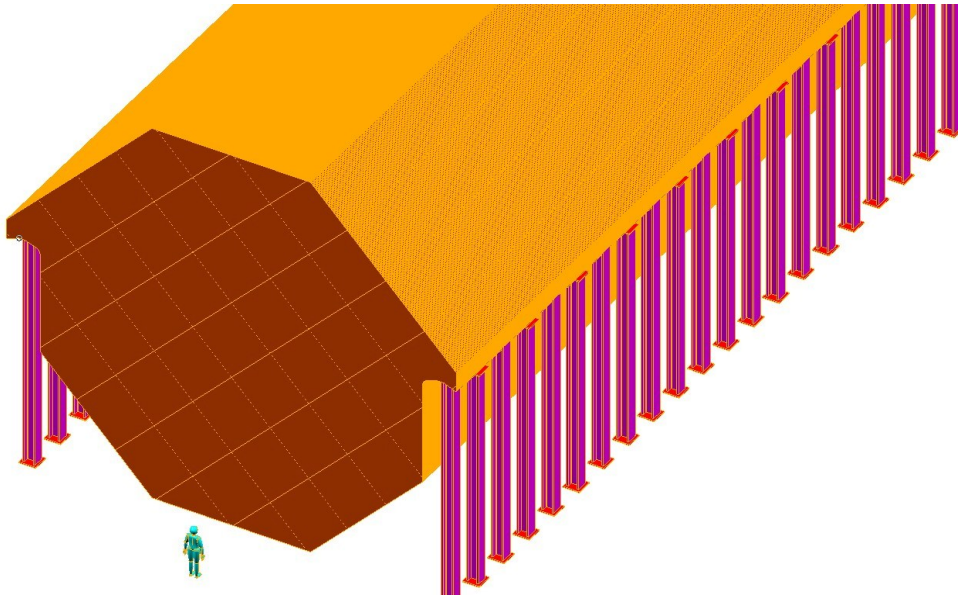

*Figure 19: The Magnetised Iron Neutrino Detector for a Neutrino Factory.*

The baseline for both the Super Beam and Beta Beam facilities is the MEMPHYS detector [8], a 500 kT fiducial mass water Cherenkov detector. This would be located in the Laboratoire Souterrain de Modane in the Fréjus tunnel in France, at a distance of 130 km from CERN. The current plan is to build the detector from two modules, 65 m in diameter and 103 m in height (see Figure 21), in two separate caverns. Based on a large experience from the SuperKamiokande experiment [32], light will be detected using 12000 8" or 10" PMTs in each module. To reduce costs, it is planned to group readout electronics [33]. To test this and other aspects of the detector, a prototype called MEMPHYNO [34] has been built at Université ParisVII and is being tested (see Figure 22). As for MIND, a simulation has been developed to determine the detector performance, also using GENIE for event generation and GEANT 4 for modeling the detector response. As an example, Figure 23 shows the reconstructed energy from identified muon rings compared with the real energy. Migration matrices have been produced for MEMPHYS and are being made available for physics performance determinations. Note that using the same detector would make it possible to run the Super Beam and Beta Beam at the same time, thereby improving the physics performance compared to both facilities alone.

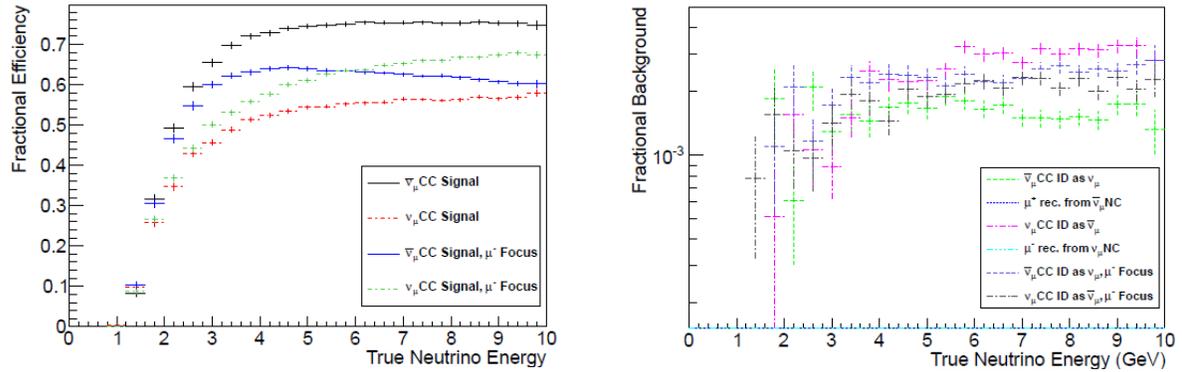

Figure 20: Performance of the MIND detector. (Left) The efficiency for detecting the muon signal events. (Right) The fractional backgrounds.

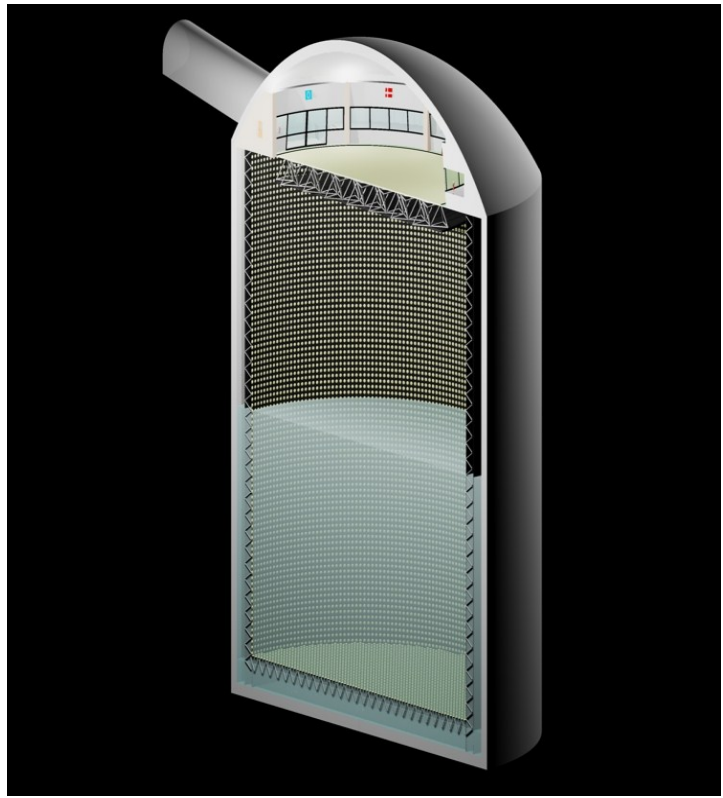

Figure 21: The proposed MEMPHYS detector for the Super Beam and the Beta Beam.

Near detectors [35] are essential for all three facilities to:
- measure the neutrino flux to 1% precision to allow the extrapolation to the far detector;
- measure the $\nu_e$ and $\nu_\mu$ cross-sections to control systematic errors;
- measure the charm production for the Neutrino Factory, as this is an important background.

In addition, the near detectors can also be used for physics, in particular the measurements of parton density functions, $\sin^2\theta_W$ and non-standard interactions from taus. A sketch of the near detector for a Neutrino Factory is shown in Figure 24. It consists of a high resolution section using a scintillating fibre tracker for flux measurements, a Mini-MIND detector for flux and muon measurements and a vertex detector for charm and tau measurements. The near detector for a Super Beam and Beta Beam would be similar, except without the vertex detector and including a water target.

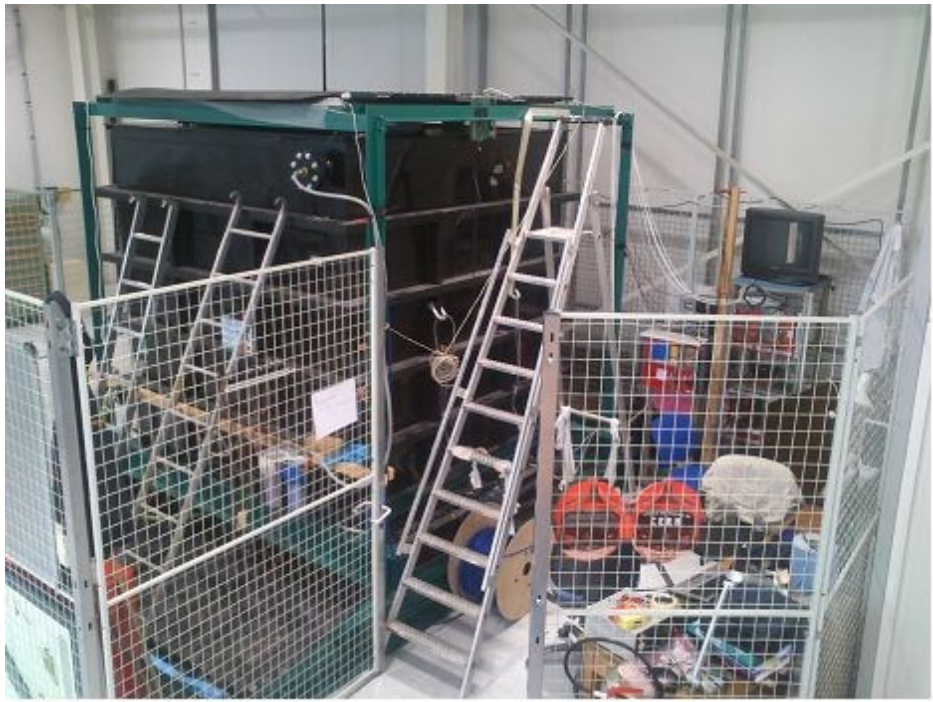

*Figure 22: The MEMPHYNO detector under test.*

**Physics Performance**

The physics group in EUROnu has determined the physics reach of each facility and combination of facilities using the parameters provided for the accelerators and detectors [1]. They have assessed and included the corresponding systematic errors in a uniform way and optimized the performance based on information from other experiments. Following the recent indications of large $\theta_{13}$, an initial physics reach comparison between the three EUROnu facilities and others has been made. The results are shown in Figures 25 to 27. For the 10 GeV Neutrino Factory (labeled LENF), the total signal systematic error used is 2.4%, while it is 5% for the other facilities. The systematic error used for the background in all cases is 10% and 10 years running time is assumed.

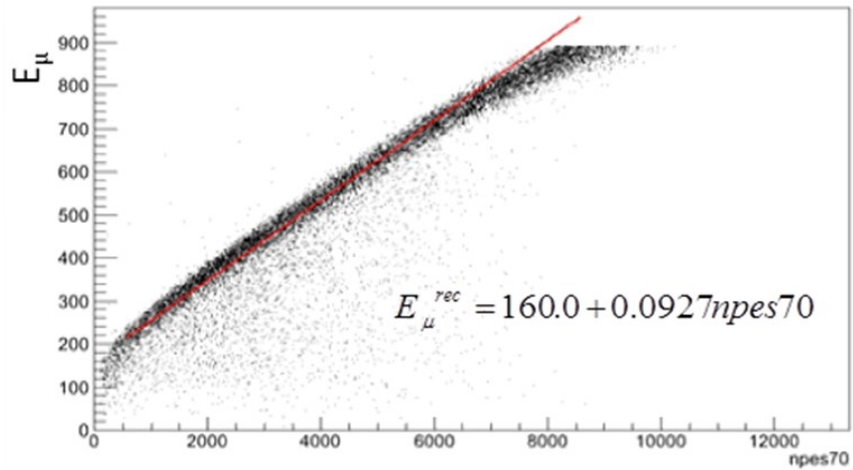

*Figure 23: Performance of the MEMPHYS detector. The reconstructed energy of a muon from a neutrino interaction is compared with the real energy, as a function of the number of photoelectrons detected.*

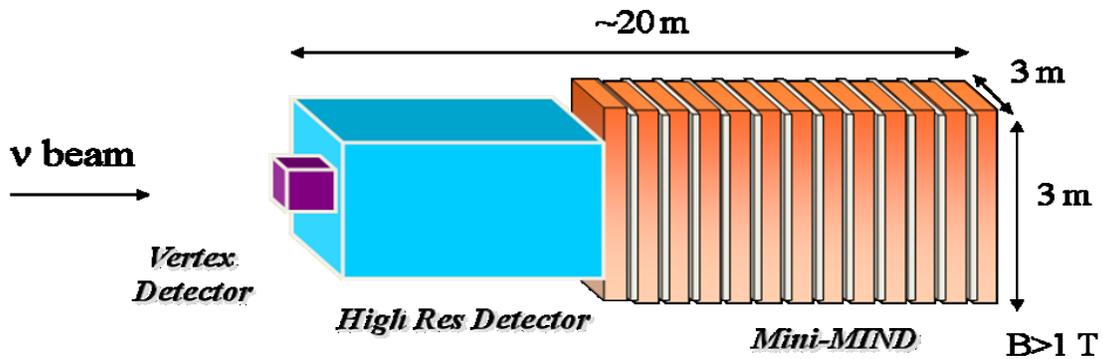

*Figure 24: The near detector for a Neutrino Factory.*

The figures clearly demonstrate that the Neutrino Factory has the best physics reach of all the future proposed projects, covering more than 80% of δ after 10 years of running and determining the mass hierarchy at 5σ on a much shorter time scale. A combination of the Super Beam and Beta Beam has the next best performance as far as the measurement of CP-violation is concerned, measuring it over 65% of δ. However, the SPL Super Beam with a detector at the second oscillation maximum, for example at the Canfranc Laboratory [11] at a distance of 630 km from CERN rather than at Fréjus, also has a very good physics reach. It has almost the same coverage of δ as the Super Beam and Beta Beam combination and could measure the mass hierarchy at 5σ for all values of δ in 10 years. Although this option has not been studied in EUROnu, it looks possible with the appropriate change in direction and downward angle of the beam.

**Costing**

As well as determining the physics performance of the three facilities, EUROnu has undertaken a costing for the construction of each. As the resources available to do this have been limited, the focus has been more on the relative cost of each facility. A lot of care has been taken to ensure similar assumptions have been made and common costs used wherever possible. For the purpose of this comparison, it has been assumed that all three facilities would be located at CERN, to put the costing on the same basis. To do this, layouts of each facility have been made on the CERN site.

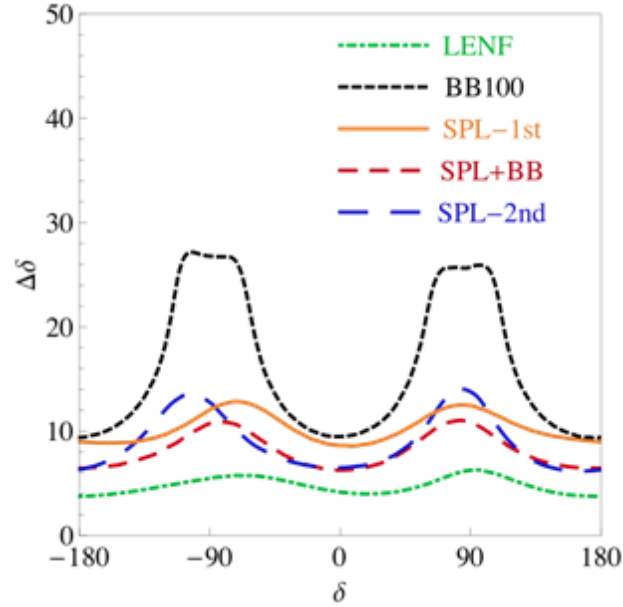

*Figure 25: The 1σ measurement errors for the CP angle δ as a function of δ. The facilities studied are as follows. LENF: the Low Energy Neutrino Factory, with a 10 GeV muon energy, $1.4 \times 10^{21}$ decays per year and a single 100 kt mass Magnetised Iron Neutrino Detector (MIND) at a baseline of 2000 km; BB100: a γ=100 Beta Beam, with $1.3/3.5 \times 10^{18}$ decays per year of Ne/He and a 500 kt Water Cherenkov detector (MEMPHYS) at Fréjus; SPL-1st: a 4 MW SPL Super Beam with 500 kt water Cherenkov detector at Fréjus, corresponding approximately to the first oscillation maximum; SPL-2nd: as above, but with the detector at Canfranc, corresponding to approximately the second oscillation maximum; SPL+BB: the combination of BB100 and SPL-1st.*

To ensure that all methodology used in the costing and all the assumptions made are well documented, a separate "Costing Paper" has been written [36]. It is essential that this document is consulted before the costs given here are used. The results of the costing are shown in Tables 1-3, taken directly from the Costing Paper. The cost is given as a lower bound and an upper bound. The lower bound is the estimated total cost, including staff costs. For each estimated cost that goes into this total, an error is also determined to reflect the uncertainties in that cost. The total error is taken to be the sum of all these errors, as this is the most conservative, though pessimistic, approach. The upper bound given is the lower bound plus this total error. Table 1 gives the estimated total cost for each of the accelerator facilities, Table 2 the estimated costs for the corresponding detectors and Table 3 the total estimated costs of the accelerator facilities and detectors.

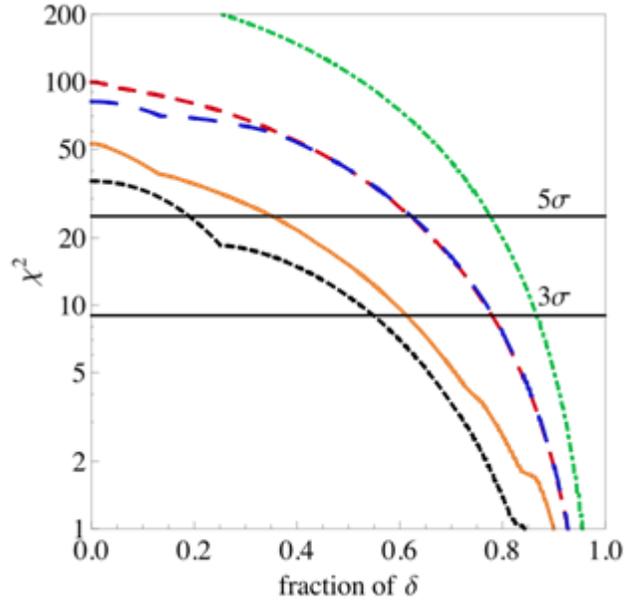

*Figure 26: The range of δ for which a 3 and 5σ measurement of CP violation can be made by the same facilities as in Figure 25.*

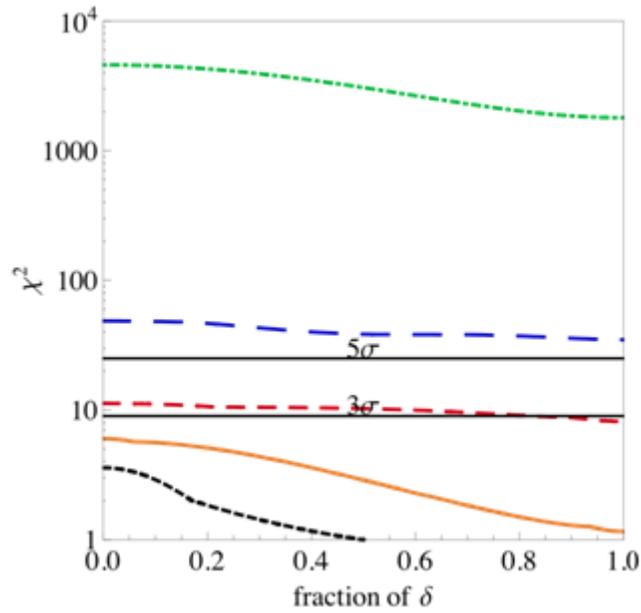

*Figure 27: The range of δ for which a 3 and 5σ measurement of mass hierarchy can be made by for the same facilities as in Figure 25.*

Table 1: Total cost of the three accelerator facilities

|  | Lower bound [MEUR] | Upper bound [MEUR] |
| --- | --- | --- |
| Super Beam | 1,193 | 1,566 |
| Beta Beam | 1,415 | 2,270 |
| Neutrino Factory | 4,663 | 6,504 |

Table 2: Total cost of the near and far detectors. The near and far detectors are the same for the Super-Beam and Beta Beam. If both facilities operated simultaneously, two near detectors would be required, but only one far detector. The near detector cost for a Neutrino Factory is for two detectors.

|  | Near detector(s) cost [MEUR] | | Far detector cost [MEUR] | | Total detector cost [MEUR] | |
| --- | --- | --- | --- | --- | --- | --- |
|  | Lower bound | Upper bound | Lower bound | Upper bound | Lower bound | Upper bound |
| Super Beam or Beta Beam | 35 | 46 | 739 | 887 | 774 | 933 |
| Neutrino Factory | 82 | 106 | 522 | 678 | 604 | 784 |

Table 3: Total cost for the accelerator facilities and the relevant detectors. Note that the lower bound without staff costs just uses a 40% scaling factor.

|  | Lower Bound [MEUR] (excluding staff costs) | Lower bound [MEUR] | Upper bound [MEUR] |
| --- | --- | --- | --- |
| Super Beam | 1,405 | 1,967 | 2,499 |
| Beta Beam | 1,564 | 2,189 | 3,203 |
| Neutrino Factory | 3,762 | 5,267 | 7,288 |

**Conclusions**

The primary aims of EUROnu have been to produce conceptual designs of a CERN to Fréjus Super Beam, a Neutrino Factory and Beta Beam and to determine their physics reach and costs. This information has then been used for a comparison between the facilities and to make a recommendation to the CERN Council on which to take forward. Based on the work done, a 10 GeV Neutrino Factory clearly has the best physics performance, but at a higher cost. It has been judged within the consortium that the physics performance offsets the additional cost and hence EUROnu has recommended the construction and operation of a 10 GeV Neutrino Factory as soon as possible [37]. To mitigate the cost, it is recommending that this is done via a staged approach, as follows:

1) Completion of the necessary design and R&D work to allow a full proposal for a Neutrino Factory to be written in 2017.
2) The construction of νSTORM [38]. This project will use an existing proton driver of around 300 kW beam power to create pions in a target. Forward going pions with an energy of 5 GeV (±10%) will be focussed into a transport line, before injection into a straight of a storage ring. Muons of around 3.8 GeV from the decay will then be transported around the ring and the neutrinos from their decay used for the following studies:
   - the search for sterile neutrinos,
   - the measurement of $\nu_e N$ scattering cross-sections,
   - neutrino detector development.
   
   In addition, this facility will be a valuable prototype for the Neutrino Factory construction.
3) The construction of a low power version of the Neutrino Factory, using an existing proton driver, without muon cooling and using a lower mass MIND detector, around 20kt. This will already have a very competitive physics potential (see Figure 28) [39].
4) The construction of the 4 MW Neutrino Factory using 10 GeV muons and a 100 kt MIND detector at a baseline of around 2000 km.

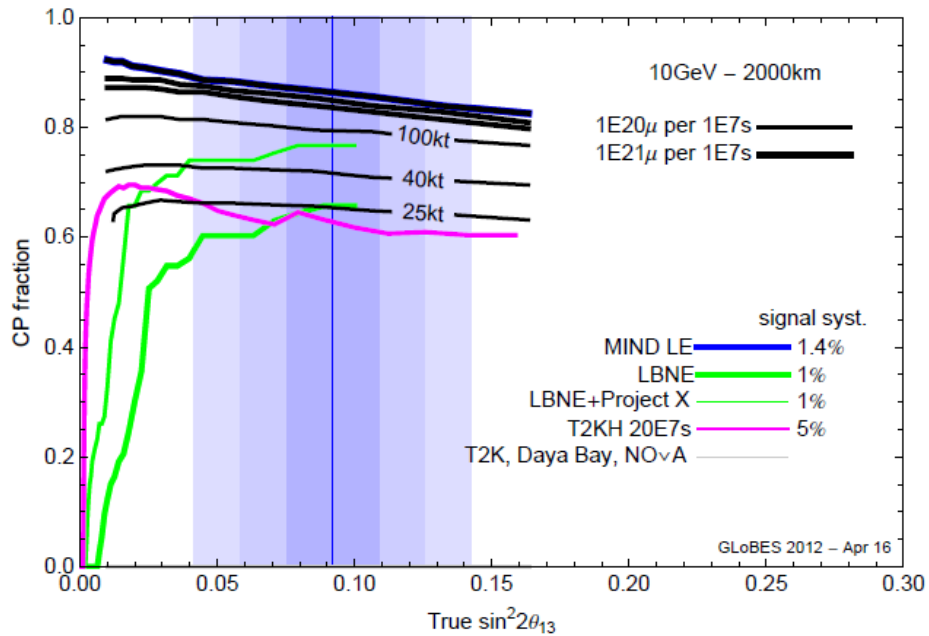

*Figure 28: A study of possible staging scenarios for a Neutrino Factory. The plot shows the fraction of the CP angle for which a measurement can be made for various MIND detector masses and the number of neutrinos produced per year. The current measured value of $\theta_{13}$ is shown as a vertical line. The potential of the Neutrino Factory is compared to other possible future facilities in the US (LBNE) and Japan (T2KH). The plot shows that a Neutrino Factory with a factor of 10 smaller neutrino flux than the full 4 MW version and a 25 kt MIND is already competitive.*

However, the SPL Super Beam is also competitive, particularly if the detector is at the second oscillation maximum. Further, much of the accelerator infrastructure for this is the same as for a Neutrino Factory. As a result, if the MEMPHYS detector was built in an underground laboratory for other physics reasons (for example, astroparticle physics and proton lifetime measurements), the additional cost of building a Super-Beam as another stage in the construction of a Neutrino Factory would be small.

This recommendation has been submitted to CERN Council via the Update of the CERN Strategy for Particle Physics 2011-2012 [40].

### Acknowledgements

We acknowledge the financial support of the European Community under the European Commission Framework Program 7 Design Study: EUROnu, Project Number 212372 and from the National Science Fund of Bulgaria under Contract number DO 02-149/07.10.2009. Neither the EC nor the Fund are liable for any use that may be made of the information herein.